# Hierarchical Machine Learning Classification of Parkinsonian Disorders using Saccadic Eye Movements: A Development and Validation Study

**SB Patel**
Nuffield Department of Clinical Neuroscience
University of Oxford
salil.patel@ndcn.ox.ac.uk

**OB Bredemeyer**
Nuffield Department of Clinical Neuroscience
University of Oxford
oliver.bredemeyer@gmail.com

**JJ FitzGerald**
Nuffield Department of Surgery
University of Oxford
james.fitzgerald@nds.ox.ac.uk

**CA Antoniades**
Nuffield Department of Clinical Neuroscience
University of Oxford
chrystalina.antoniades@ndcn.ox.ac.uk

## Abstract

Discriminating between Parkinson's Disease (PD) and Progressive Supranuclear Palsy (PSP) is difficult due to overlapping symptoms, especially early on. Saccades (rapid conjugate eye movements between fixation points) are affected by both diseases but conventional saccade analyses exhibit group level differences only. We hypothesised analysing entire saccade raw time series waveforms would permit superior individual level discrimination between PD, PSP, and healthy controls (HC). 13,309 saccadic eye movements from 127 participants were analysed using a novel, calibration-free waveform analysis and hierarchical machine learning framework. Individual saccades were classified based on which trained model could reconstruct each waveform with minimum error, indicating the most likely condition. A hierarchical classifier then predicted overall status (recently diagnosed and medication-naive 'de novo' PD, 'established' PD on antiparkinsonian medication, PSP, and healthy controls) by combining each participant's saccade results. This approach substantially outperformed conventional metrics, achieving high AUROCs distinguishing de novo PD from PSP (0.92-0.97), de novo PD from HC (0.72-0.89), and PSP from HC (0.90-0.95), while the conventional model showed limited performance (AUROC range: 0.45-0.75). This calibration-free waveform analysis sets a new standard for precise saccadic classification of PD, PSP, and HC, increasing potential for clinical adoption, remote monitoring, and screening.

## Introduction

Parkinson's Disease (PD) and Progressive Supranuclear Palsy (PSP) are distinct neurodegenerative disorders with entirely different pathophysiological origins. PD, a synucleinopathy, is characterized principally by the degeneration of dopaminergic neurons in the substantia nigra pars compacta, resulting in motor symptoms such as resting tremor, bradykinesia, rigidity, and postural instability. [1,2] In contrast, PSP is a tauopathy, distinguished by the accumulation of abnormally hyperphosphorylated tau protein in subcortical and brainstem nuclei. Despite the different pathology, this at first leads to symptoms that resemble PD, leading to PSP being classed as an 'atypical parkinsonian syndrome'. [3,4] Other features later supervene, including progressive postural instability, vertical supranuclear gaze palsy, and cognitive dysfunction. [5,6]

The overlapping clinical presentations of PD and PSP, particularly in the early stages, pose significant diagnostic challenges, often leading to misdiagnosis. Accurate differentiation is crucial for prognostication and optimizing treatment strategies, as PSP responds poorly to levodopa, the first-line treatment for PD. [7] Despite advances in understanding these conditions, current diagnostic approaches rely largely on subjective clinical assessment, while objective biomarkers for their diagnosis during life have remained elusive. [8-10]

Saccadic eye movements have emerged as a promising line of investigation. [11-13] Saccades are rapid, conjugate eye movements from one fixation point to another, critical for normal visual perception and cognition.[14] The generation and control of saccades relies on neurocircuitry involving the brainstem, superior colliculus, cerebellum, and various cortical areas including the frontal eye fields.

In PD, the degradation of dopaminergic substantia nigra neurons disrupts this balanced control, leading to clinical manifestations such as increased saccadic latency and hypometric trajectories.[15,16] In contrast, PSP patients demonstrate more pronounced oculomotor dysfunction, with slowed saccade velocity, increased latency, impaired saccade accuracy, and disinhibition.[17,18]

Conventional metrics of saccades such as amplitudes and velocities are seen to differ between PD, PSP, and healthy controls at the group level, however the distributions overlap and so they are of little diagnostic use for classification at the level of the individual patient. [19-21] These metrics capture only a fraction of the information present in the full saccadic waveform, and we hypothesise that analysis based on the entire waveform would perform better.

Clinical use of saccadic eye movements poses significant additional challenges.[22] Calibration of eye trackers requires frequent adjustment and close participant cooperation, which can be difficult with some patient groups. Instrumental inaccuracies and calibration difficulties introduce noise and variability, complicating differentiation of pathological changes from normal variability. [23,24] Extracting precise features from raw data is also complicated by fixation misclassifications and over-segmentation. [25,26] These issues necessitate precise preprocessing and procedural standardisation.

To address these limitations, we introduce a novel waveform reconstruction method that utilises machine learning models in series to assess the similarity between eye movements from a given patient, and those from patients with a known diagnosis, to predict the patient's disease status. We compare several variations of this reconstruction approach, using established machine learning techniques including PCA reconstruction, naive Bayes classifiers, and deep learning models such as convolutional autoencoders and multi-layer perceptron (MLP) classifiers, against a machine learning approach based on conventional metrics. [27-30] Importantly, our method does not require calibration of the eye-tracking device, reducing task complexity and patient participation demands, thereby mitigating challenges associated with frequent adjustments and the need for patient cooperation. We show that analysing the complete raw waveform in this way both dramatically improves the performance of saccade analysis as a biomarker, and simultaneously minimises the need for extensive pre-processing and calibration, addressing concerns related to noise and variability inherent in conventional approaches.

## Results

| | "De novo", medication naive PD | Established, medicated PD | PSP | Healthy Controls |
|---|---|---|---|---|
| Mean time since diagnosis from now, months (SD) | 23.14 (26.62) | 56.82 (51.03) | 21.67 (20.57) | n/a |
| Participants | 11 | 69 | 12 | 35 |
| Number of Saccades | 1268 | 7171 | 916 | 3954 |
| Saccades per participant | 115.27 | 103.93 | 76.33 | 112.97 |
| Age at time of testing, years (SD) | 65.45 (5.13) | 65.67 (7.34) | 69.75 (10.32) | 67.66 (6.39) |
| Female | 6 (54.55%) | 26 (37.68%) | 5 (41.67%) | 17 (48.57%) |
| Male | 5 (45.45%) | 43 (62.32%) | 7 (58.33%) | 18 (51.43%) |

Table 1: *Participant characteristics by group.*

13,309 saccades, from 127 participants, were analysed (Table 1). The conventional RF+NB model, based on conventional saccade metrics, showed limited performance across most pairwise comparisons. While it achieved an AUROC of 0.75 for distinguishing de novo PD from PSP and PSP from healthy controls, it performed poorly in differentiating de novo PD from established PD (AUROC: 0.45) and showed only moderate accuracy for other comparisons (AUROC range: 0.60-0.64).

In contrast, the waveform reconstruction methods demonstrated substantially improved classification performance across most tasks. The PCA-based models (PCA+NB and PCA+MLP) consistently exhibited high accuracy. The PCA+NB model achieved an AUROC of 0.94 for distinguishing de novo PD from PSP, 0.80 for established PD vs. PSP, and 0.89 for de novo PD vs. healthy controls. The PCA+MLP model showed even better performance in some tasks, with AUROC scores of 0.97, 0.86, and 0.72 for the same comparisons, respectively.

The Sym_AE-based models (Sym_AE+NB and Sym_AE+MLP) also demonstrated strong performance, particularly in distinguishing de novo PD from PSP (AUROC: 0.92-0.93) and established PD from PSP (AUROC: 0.82-0.85). However, their accuracy was lower when comparing de novo PD or established PD to healthy controls (AUROC range: 0.53-0.66).

Notably, all waveform reconstruction models achieved high accuracy in distinguishing PSP from healthy controls, with AUROC scores between 0.90 and 0.95. This finding suggests that these methods could potentially aid in the early detection of PSP.

Furthermore, the waveform reconstruction models were able to differentiate between de novo and established PD patients, with the PCA+NB model achieving an AUROC of 0.89 and the Sym_AE+NB model achieving an AUROC of 0.81. Given that these two groups differed primarily by medication status and disease duration, these results suggest that the waveform approach can potentially uncover previously undetected effects of dopaminergic therapy on saccade trajectories.

Overall, the PCA+NB model emerged as the best-performing waveform reconstruction approach, consistently achieving high AUROC scores across most pairwise comparisons. It demonstrated excellent performance in distinguishing de novo PD from PSP (AUROC: 0.94), established PD from PSP (AUROC: 0.80), and de novo PD from healthy controls (AUROC: 0.89). Additionally, it achieved the highest accuracy among all models in differentiating PSP from healthy controls (AUROC: 0.95) and de novo PD from established PD (AUROC: 0.89).

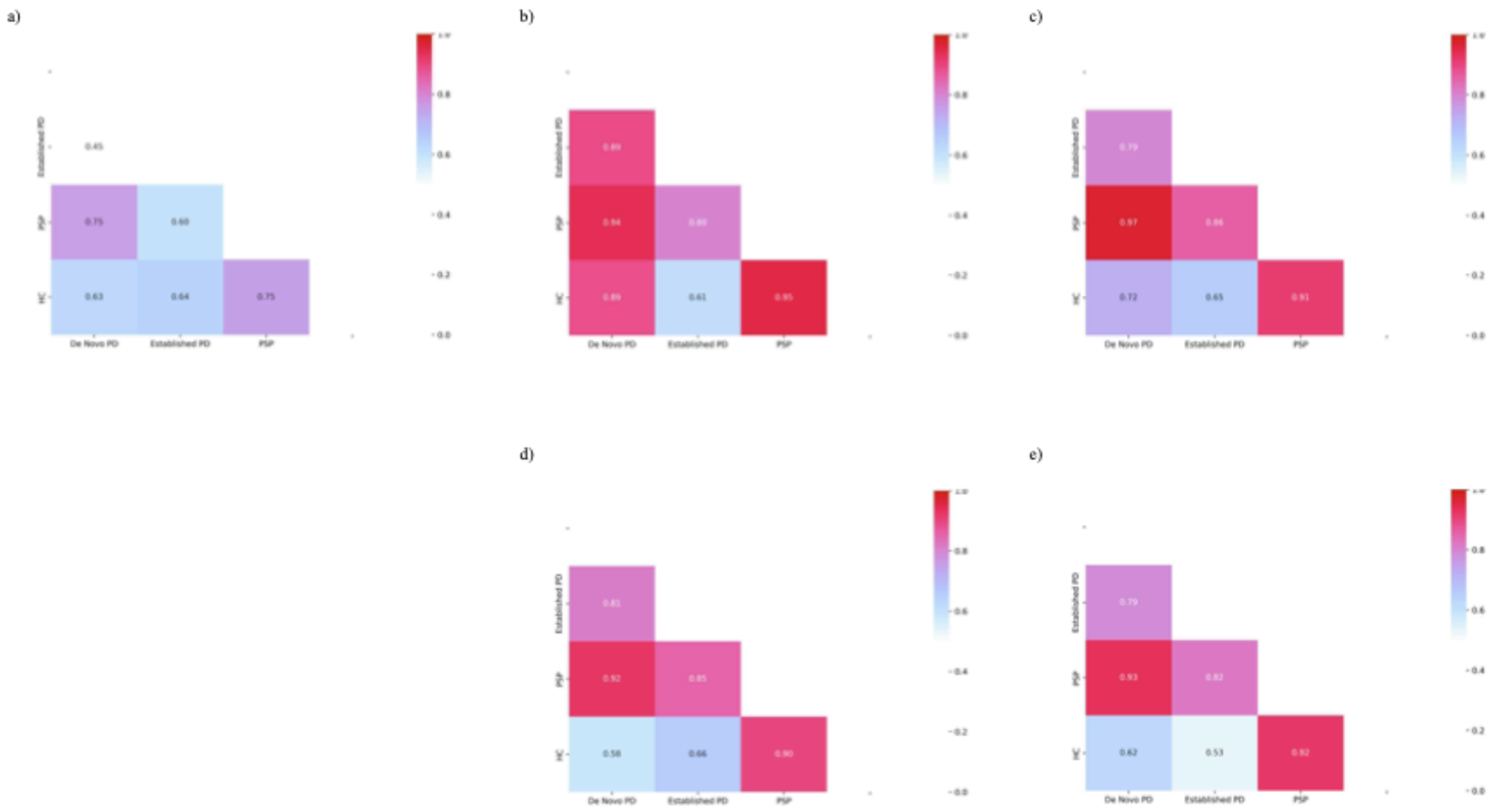

Figure 1: *Heatmaps showing the mean area under the receiver operating characteristic curve (AUROC) scores for pairwise classification of participant groups using (a) conventional metrics with Random Forest followed by Gaussian Naïve Bayes, (b) principal component analysis with Gaussian Naïve Bayes (PCAxNB), (c) principal component analysis with multilayer perceptron (PCAxMLP), (d) symmetric autoencoder with Gaussian Naïve Bayes (Sym_AExNB), and (e) symmetric autoencoder with multilayer perceptron (Sym_AExMLP). AUROC scores were calculated using quicksort leave-pair-out cross-validation.*

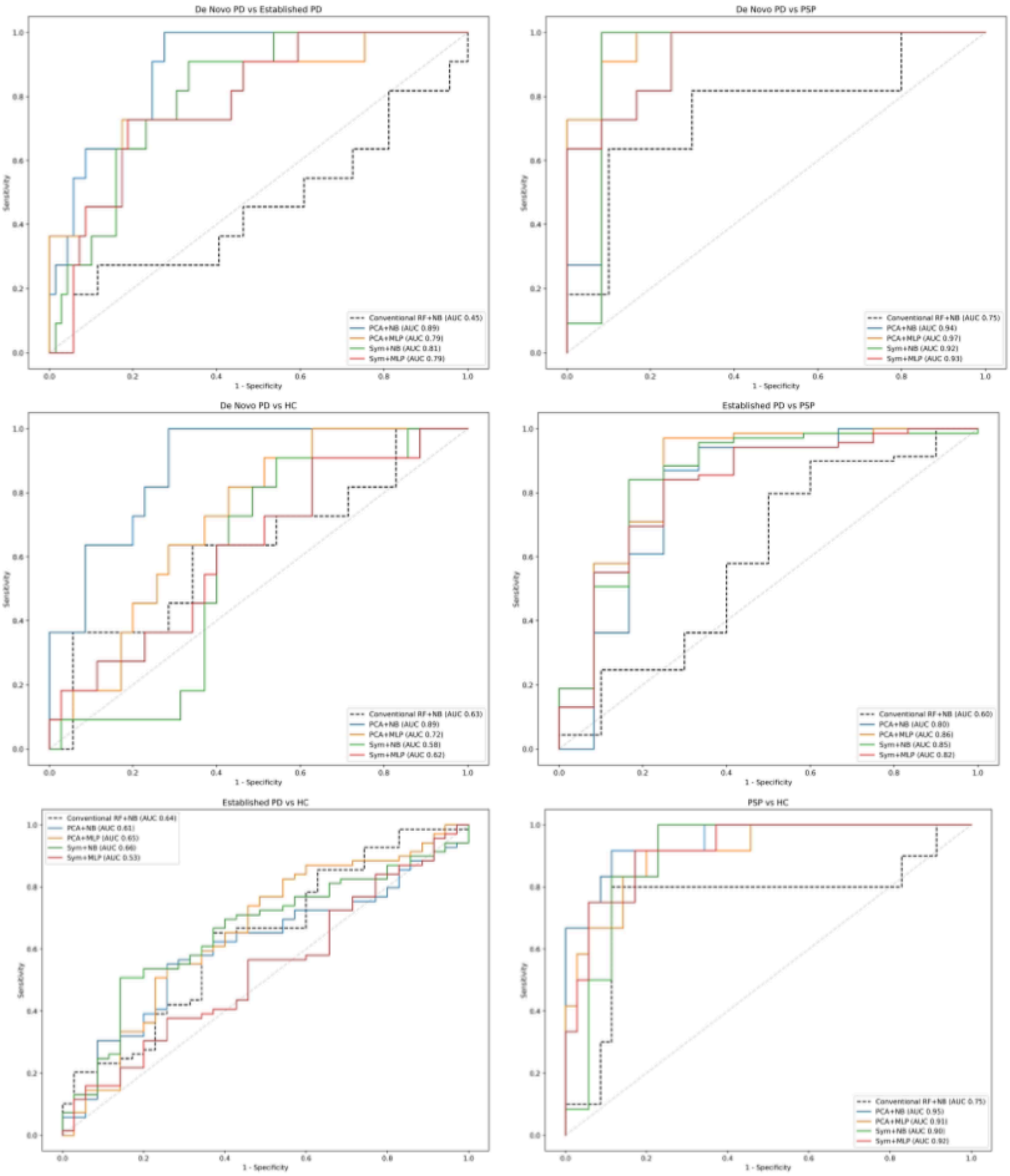

Figure 2: *Receiver operating characteristic (ROC) curves comparing the performance of waveform reconstruction models (colored lines) against the conventional metric model (black dashed line) for pairwise classification between the four participant groups. The conventional metric model used Random Forest followed by Gaussian Naïve Bayes, while the waveform models included PCAxNB, PCAxMLP, Sym_AExNB, and Sym_AExMLP. The area under each ROC curve (AUROC) was calculated using quicksort leave-pair-out cross-validation to assess model performance.*

**Discussion**

Assessments of saccadic eye movements have traditionally focused on metrics such as latency, velocity, amplitude, and more recently the damping ratio. While group level abnormalities of these measures are well documented in PD and other parkinsonian like disorders such as PSP, their ability to predict disease state at the individual level is very limited. In this study, we have introduced two innovations. First, analysis of the entire saccadic waveform. Notably, analysis of the velocity waveform consistently outperformed the position waveform across the waveform reconstruction models (Supplementary Material 4). Second, the use of a hierarchical ML approach that first analyses the individual saccades for characteristics suggestive of a particular condition, then looks at the results of this across all saccades for an individual to determine the most likely diagnosis from the entire collection.

Both components are necessary for optimal performance. When the hierarchical system was used with conventional metrics rather than waveforms as input into the first stage, it had low predictive utility, illustrating the importance of the whole waveform analysis. The waveform reconstruction approach outperformed this conventional metric analysis across all diagnostic comparisons, with the single exception of differentiating individuals with established PD from control participants. The improved performance overall was despite the new system being blind to calibration-dependent information such as amplitude - traditionally considered important in quantifying parkinsonian disorders.[31]

High classification scores were achieved when comparing de novo PD patients to healthy controls strongly suggesting that this method will be valuable in early diagnosis of PD. The hierarchical waveform reconstruction approach also shows significant promise in differentiating between idiopathic PD and PSP, currently a major diagnostic challenge. Specialists are often not able to accurately distinguish PSP from PD due to phenotypic overlap at presentation, and the correct diagnosis is only made with the passage of time, failure of dopaminergic treatments, and emergence of later features of PSP. Saccade generation requires complex coordination of frontal cortices, basal ganglia, superior colliculus, cerebellar vermis and brainstem nuclei.[32] The ability of the waveform reconstruction approach to separate these two conditions presumably reflects differences in how the two pathologies affect components of this network, leading to distinct aberrances of saccadic waveforms.

The difficulty in distinguishing between established PD and healthy controls across all models may be attributed to the masking effect of dopaminergic treatments.[33] Dopaminergic medications, such as levodopa, are used to alleviate motor symptoms in PD patients, and in doing so they normalise many of the somatomotor signs upon which clinical assessment is made. It is not surprising that they similarly make the saccadic eye movements of individuals with PD more like those of healthy controls. This finding highlights the importance of considering medication status when interpreting saccadic eye movement data in patients.

A significant advantage of the methodology used in this study is the avoidance of the need for calibration. The approach focuses on the morphology of the waveform and is invariant to its the absolute position and amplitude. The removal of the need for calibration will increase ease of use for non-specialists, and therefore accessibility to patients. It may provide a tool to guide primary care physicians as to when referral to a specialist is appropriate and facilitate earlier diagnosis and timely treatment. It expands the potential application of our approach beyond clinical settings, facilitating remote monitoring and possibly even community-based screening. By accurately distinguishing between de novo PD, PSP, and healthy controls, our method could enable early detection and intervention, ultimately leading to improved patient outcomes.

The sample size for the PSP group was smaller than the PD or control groups due to the rarity of the condition, and PSP patients on average completed fewer saccades compared to other groups (Table 1), mainly due to fatigue. The reconstruction error classifier and cross-validation technique are designed to be robust to class imbalance, but nevertheless, future studies should aim to validate our findings in larger groups to ensure the accuracy of our approach and provide external validation with independent datasets from different centres.

The proliferation of technology capable of eye tracking, such as smartphones, tablets and augmented reality headsets, may facilitate widespread use of a system like this. However, this approach needs to be validated using additional devices; while it is calibration-independent, the effects of device-specific linearity errors and spatiotemporal resolution should be investigated. In particular, the richness of information contained in the saccadic waveform will decrease rapidly with falling sampling rate (Supplementary Material 5). Few consumer devices available at present can match the saccadometer's sampling rate, though the cameras on such devices are likely to improve rapidly.

In summary, our study presents a novel, scalable, and non-invasive method for distinguishing between de novo and established PD, PSP, and healthy control groups using waveform reconstruction techniques. The high discriminative capability of our approach, combined with its calibration-free nature and potential for widespread clinical adoption, offers a promising path towards improved early diagnosis, treatment monitoring, and improved patient outcomes in Parkinsonian disorders. While future work should focus on external validation and the integration of our method with other biomarkers, the results presented here highlight the significant potential of saccadic eye movement analysis as an objective biomarker in the clinical management of Parkinsonian disorders. By refining and expanding upon this approach, we can work towards a new standard of care in the diagnosis and treatment of these complex neurodegenerative conditions.

## Methods

*Participant Recruitment*

Participants were recruited as part of the OxQUIP study, which included individuals diagnosed with Parkinson's Disease (PD), Progressive Supranuclear Palsy (PSP), and healthy controls (HC). Inclusion in the PD and PSP groups required a diagnosis made by a neurologist according to the UK Brain Bank criteria, while age-matched healthy controls were selected based on the absence of any known neurological disorders. [34] Controls were often patients' partners, to control for potential socioeconomic confounding factors. Patients receiving deep brain stimulation to treat PD were excluded. An ethics committee (Research Ethics Committee reference 16/SW/0262) approved the study, and experiments were conducted in accordance with ICH Good Clinical Practice guidelines and the Declaration of Helsinki.

The PD cohort was divided into two subgroups: 'de novo' PD patients who were recently diagnosed and medication naive, and an "established" PD patient who were on antiparkinsonian medication. Together with PSP and HC groups, there were four groups for analysis.

All saccadic eye movement data were collected using a single infrared oculometer (Saccadometer Advanced; Ober Consulting, Poznán, Poland). Saccadometry was conducted in accordance with a previously published internationally standardised protocol.[35] Participants were seated in a dimly lit room and instructed to fixate on a centrally located visual target. The prosaccade (PS) task required participants to shift their gaze towards a suddenly appearing peripheral target, 10 degrees horizontally to either the right or left of the central fixation point. During each trial, the saccadometer detected saccadic responses online and captured the onset, velocity, trajectory, and termination of each observed saccade. Eye position during the saccade was sampled at 200 Hz.[36] Each session included 120 saccade trials in two blocks of 60. However, some saccades were not successfully recorded due to both technical factors (such as calibration issues, signal noise, or data acquisition problems) and non-technical factors (such as participant fatigue, blinking, or head movements). The mean number successfully recorded was 105 per participant (though this differed between groups; Table 1).
Saccadic data were processed using Python (version 3.10) with open-source Pandas, NumPy, and SciPy packages.

Hierarchical Classification

We adopted a hierarchical classification strategy, using two machine learning models in series to form a prediction. The first model forms predictions at the level of an individual saccade, as well as optionally encoding a low-dimensionality representation of the input data. Predictions and encodings are then aggregated over all saccades made by an individual by drawing evenly spaced quantiles from the distributions of the predicted class probabilities and encodings of the model. This aggregation step synthesises a comprehensive feature set that captures both the central tendencies and the dispersion of the predictive signals across the dataset, which is then used as input for a second machine learning model to make predictions at the level of a participant (Figure 1)[37]. This hierarchical approach allows us to use all available data points for each participant, to capture the variability between individual saccadic trajectories, and evaluate different classification algorithms by extracting features to obtain a single prediction for a given individual.

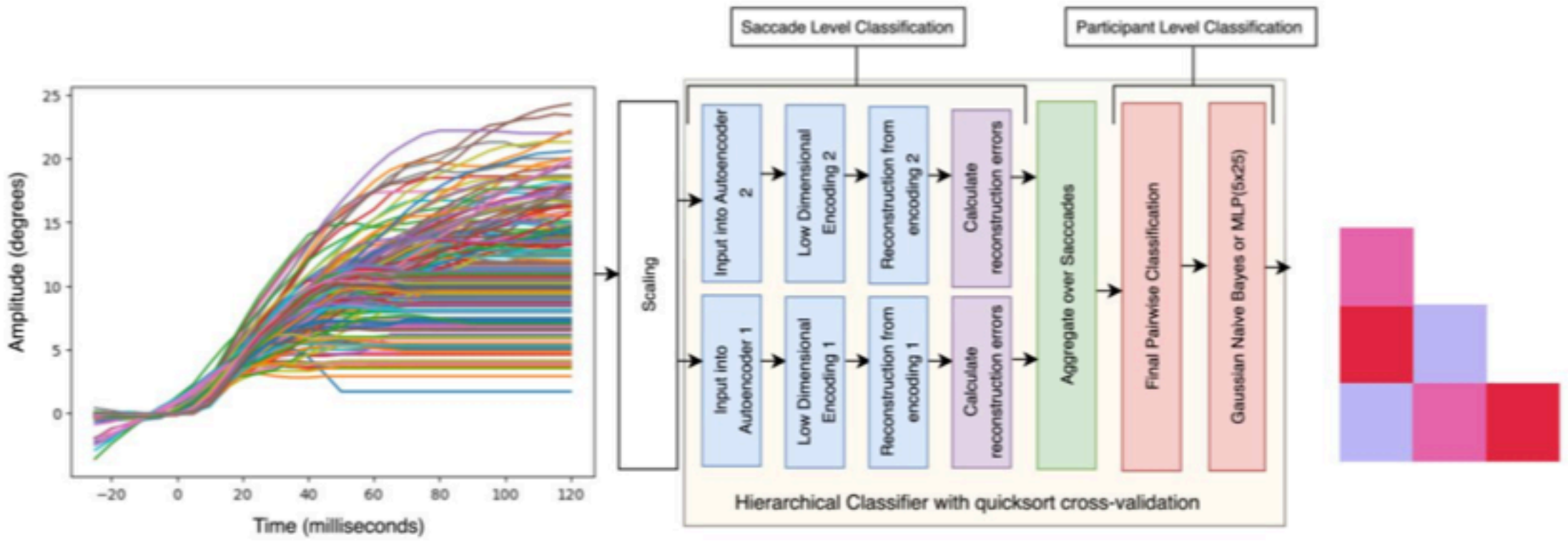

Figure 3: *The novel waveform reconstruction method consists of a hierarchical classification pipeline with dual-autoencoders, a reconstruction classifier, voting aggregation, and participant level naive Bayesian/Multi-Layer Perceptron (MLP) classifiers as components. The initial phase reconstructs and classifies individual saccades, and the final phase aggregates all individual saccades from the same participant and classifies at the participant level for an individual diagnosis. Raw saccade waveforms, here shown from a single participant, are the input and cross-validated group pairwise comparisons are shown as the output visualisation.*

*Waveform Reconstruction Method*

Previous studies have used autoencoders for anomaly detection by calculating the reconstruction error of an autoencoder trained on normative data and using the error as a measure of abnormality for the input data point.[38] Here we extend this approach by training autoencoders for each participant group independently, and using the reconstruction errors from each group-specific autoencoder in parallel to classify each saccade, hereafter referred to as a “reconstruction error classifier” (REC). A lower reconstruction error for a particular group implies a closer fit to the REC’s learned representation of that group, thereby indicating group membership. Conversely, a higher error signals a deviation from the model's representation, suggesting membership of a different group.

We explored two ways of pre-processing the raw saccadic waveforms. First, we took as input to our models the eye position over time, normalised so that the eye position furthest from the target prior to the saccade was 0 and the final eye position at the end of the saccade was 1 (“position input”). Second, we used the velocity waveform as input (calculated from the normalised position by numerical differentiation; “velocity input”).

We implemented PCA-based and deep autoencoders as dimensionality reduction techniques to capture the latent structure and dynamics of the raw saccade waveforms. [39,40] The autoencoder architecture reduces dimensionality and distils the waveform features into a more manageable representation. We tested multiple architectures, including a PCA-based model for linear dimensionality reduction (PCA_AE) and a symmetrical convolutional autoencoder (Sym_AE) designed to harness the inherent spatial hierarchies in data through convolution and pooling layers. Additionally, we explored recurrent neural networks such as Long Short-Term Memory (LSTM) and Gated Recurrent Units (GRU), as well as Temporal Convolutional Networks (TCN), and transformers to analyse the saccade data.[41] However, the results from these deeper models were suboptimal, likely due to their complexity not being well-suited to our short time series raw data. Consequently, we focus on the PCA_AE and Sym_AE models in this study.

*Conventional Metrics Classification*

For conventional metrics classification, a Random Forest (RF) classifier was employed. The model was trained using an ensemble of 400 decision trees, each developed on a bootstrap sample of the training data and incorporating random selection of features when splitting nodes to prevent overfitting. The features selected for the Random Forest model were latency, amplitude, peak velocity, mean velocity, saccade direction, whether multiple saccades were performed, peak and mean acceleration and duration of the acceleration phase, peak and mean deceleration and duration of the deceleration phase and estimates and errors of the damping ratio as well as the overall fit error. This approach builds upon our previously published methodology (though with the new inclusion of the PSP cohort and PD subgroups) introducing the hierarchical approach for aggregation and participant-level classification. [42] For the conventional metrics approach, saccades from blocks where no calibration factor could be calculated were excluded (Figure 2). The calibration process is crucial for deriving accurate saccade metrics but can be labour-intensive (Supplementary Material 1).

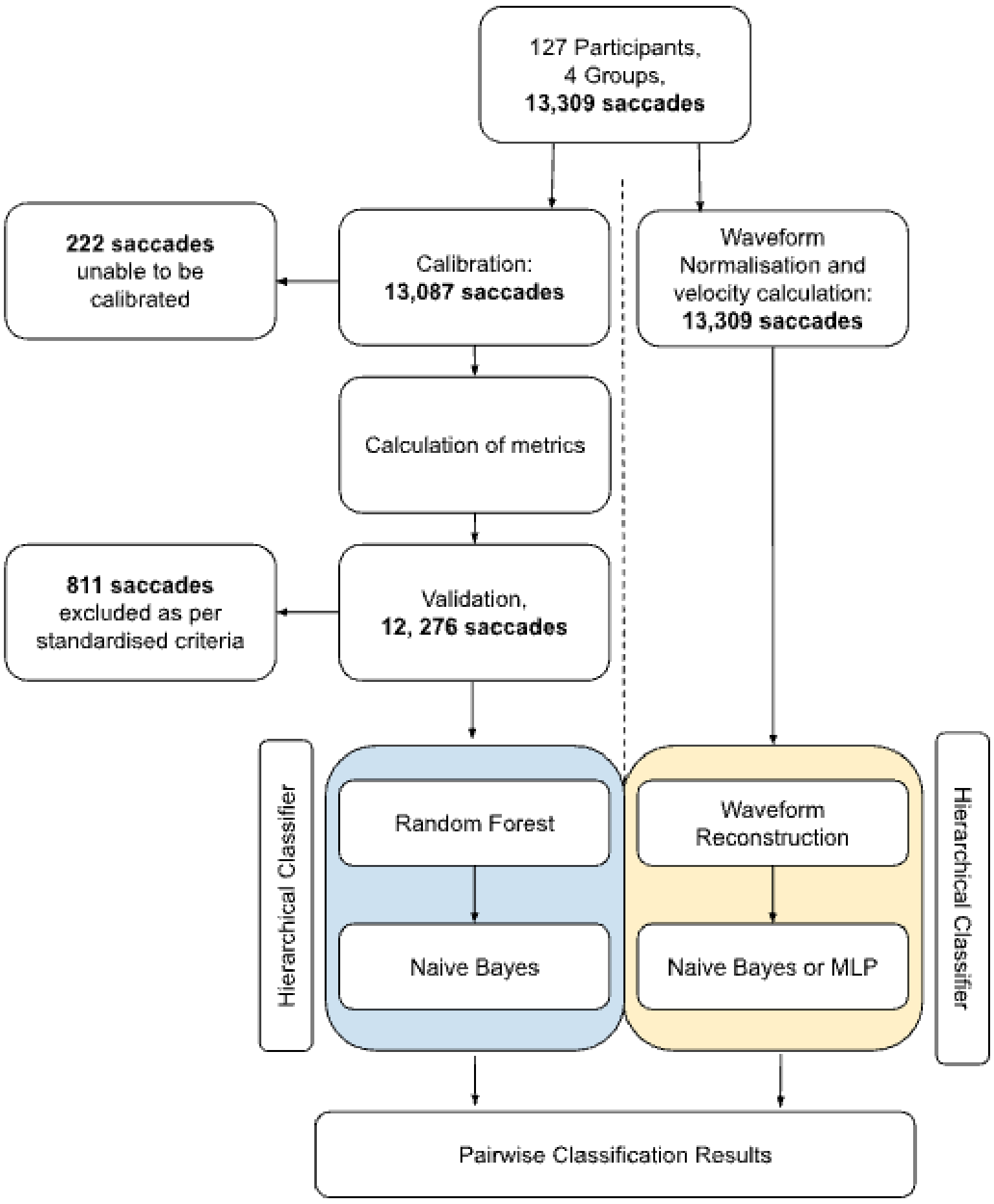

Figure 4: *The number of saccades unable to be calibrated and excluded for conventional metrics analyses versus the waveform reconstruction approach requiring no calibration.*

*Model evaluation*

The final stage of the hierarchical pipeline, for both the waveform reconstruction and the conventional metrics approach, is classification using Gaussian Naïve Bayes (NB) or a simple deep neural network (MLP; multilayer perceptron, 3 hidden layers of 25 neurons each, ReLU activation). All analyses were purposefully blinded to any clinical demographic data, such as age or gender, so performance was based on conventional metrics or waveforms alone. Each model's performance was evaluated in a binary classification task between all pairs of groups. Models were evaluated by quicksort leave-pair-out cross-validation (QLPO-CV) to generate estimates of the area under the receiver operating characteristic (AUROC) and average precision (AP). [43] QLPO-CV was chosen as it aligns with our primary objective of classification, provides unbiased estimates for group imbalanced datasets, and offers a more accurate estimate of the AUROC compared with other cross validation methods such as stratified k-fold cross-validation (Supplementary Material 2). An estimate of the optimal sensitivity and specificity of each model for our dataset was obtained from the ROC by maximising the Youden J-statistic (Supplementary Material 3). [44]

**Data Sharing**

Code and data may be available upon reasonable request and under a data sharing agreement, as permitted by our ethics committee. For formal inquiries or requests, please contact Professor Chrystalina Antoniades at chrystalina.antoniades@ndcn.ox.ac.uk.

**Competing Interests**

The authors declare no competing interests.

## Supplementary Materials

### 1. Conventional Metric Analysis and Calibration:

Calibration is a crucial step in the analysis of saccadic eye movements using conventional metrics. It ensures that the measured eye positions accurately reflect the true eye positions, enabling the precise calculation of saccade parameters such as latency, amplitude, and peak velocity. The calibration process involves participants performing fewer trials where they fixate on known target positions, establishing an accurate mapping between the recorded eye positions and the actual gaze positions.

In our study, participants performed calibration trials before each block of 60 saccadic movements. The calibration data were used to derive a calibration factor, which was then applied to the eye position data from the corresponding experimental block. This process corrects for any discrepancies between the measured and actual eye positions, ensuring the accuracy of the derived saccade metrics.

However, the calibration process can be labour-intensive and time-consuming, as it requires participants to complete additional trials and, at times, researchers to carefully inspect and validate the calibration data. In some cases, calibration may fail due to factors such as participant non-compliance, equipment issues, or data quality problems. Researchers manually inspect data and assess whether to include, with adjustments or exclude these trials. In our analysis, saccades from blocks where no calibration factor could be calculated were excluded. Out of a total of 606 blocks from 127 participants, 594 blocks were successfully calibrated, while saccades from 9 blocks were removed due to calibration issues. This results in 222 saccades being removed from the conventional metrics analysis from the following groups; 3 established PD, 4 PSP and 2 healthy controls.

After calibration, a validation step was performed to further filter saccades based on specific criteria, ensuring data quality. Saccades with latencies less than 100 ms or greater than 1,000 ms were excluded, as they likely represent anticipatory or delayed responses rather than true visually-guided saccades. Finally, saccades with peak velocities greater than 1,000°/s were excluded, as they are physiologically implausible and likely represent measurement errors. This removed 811 saccades from out conventional metrics analysis.

The conventional saccade metrics analysed in our study include latency, peak velocity, amplitude, and damping ratio. Saccadic latency, measured in milliseconds (ms), represents the time delay between the onset of a visual stimulus and the initiation of the saccadic eye movement. It reflects the time required for visual processing, decision-making, and motor planning. Peak saccadic velocity, expressed in degrees per second (°/s), is the maximum speed reached by the eye during a saccade and provides information about the dynamics of the saccadic system. Saccadic amplitude, measured in degrees (°), represents the angular distance travelled by the eye between the start and end of a saccade, reflecting the accuracy and precision of the saccadic targeting system. The damping ratio, a dimensionless value, quantifies the resistance of the saccadic system to oscillations. A lower damping ratio indicates a greater tendency for oscillations, which may be indicative of certain neurological conditions.

In contrast to the conventional metrics approach, our waveform reconstruction method relies on the morphology of the saccade waveform rather than precise eye position information. By focusing on the shape and temporal characteristics of the saccade waveform, we can circumvent the need for a labour-intensive calibration process. This approach treats the saccade waveform as a time series, where the relative changes in eye position over time are more informative than the absolute eye positions.

The waveform reconstruction method employs autoencoders to learn a compressed representation of the saccade waveforms. By training autoencoders on saccade waveforms from different participant groups, the model learns to capture the characteristic temporal patterns and morphological features specific to each group. During classification, the reconstruction error of a saccade waveform is used as a measure of its similarity to the learned group representations, allowing for the classification of saccades based on their waveform characteristics, without relying on precise eye position information. This approach effectively eliminates the need for calibration and minimizes data loss due to calibration failures, as saccades can still be analysed and classified based on their waveform morphology.

2. **Cross Validation Comparison**

We provide a direct comparison of AUC values between QLPO-CV and stratified k-fold cross-validation for all pairwise comparisons. The heatmaps shown are calculated using our best performing PCAxNB model and demonstrate the similarities in binary pairwise classification performance between the two cross-validation methods, indicating that QLPO-CV provides results consistent with the more commonly used stratified k-fold cross-validation. While stratified k-fold cross-validation can be used to estimate metrics such as accuracy and F1 score, these metrics are not directly applicable to QLPO-CV, which is specifically designed for AUROC estimation. QLPO-CV focuses on generating a ranking of samples based on their predicted scores, aiming to provide unbiased estimates of the ROC curve and AUC. This is especially useful for imbalanced datasets.

Accuracy and F1 scores are heavily influenced by class frequencies and are typically calculated on a separate validation cohort or through methods such as stratified k-fold cross-validation, which can sometimes underestimate or overestimate performance depending on dataset characteristics. By using QLPO-CV, we ensure a robust and reliable estimation of the ROAUC, aligning with our primary analysis objectives (binary pairwise classification) while maintaining consistency with established cross-validation methods.

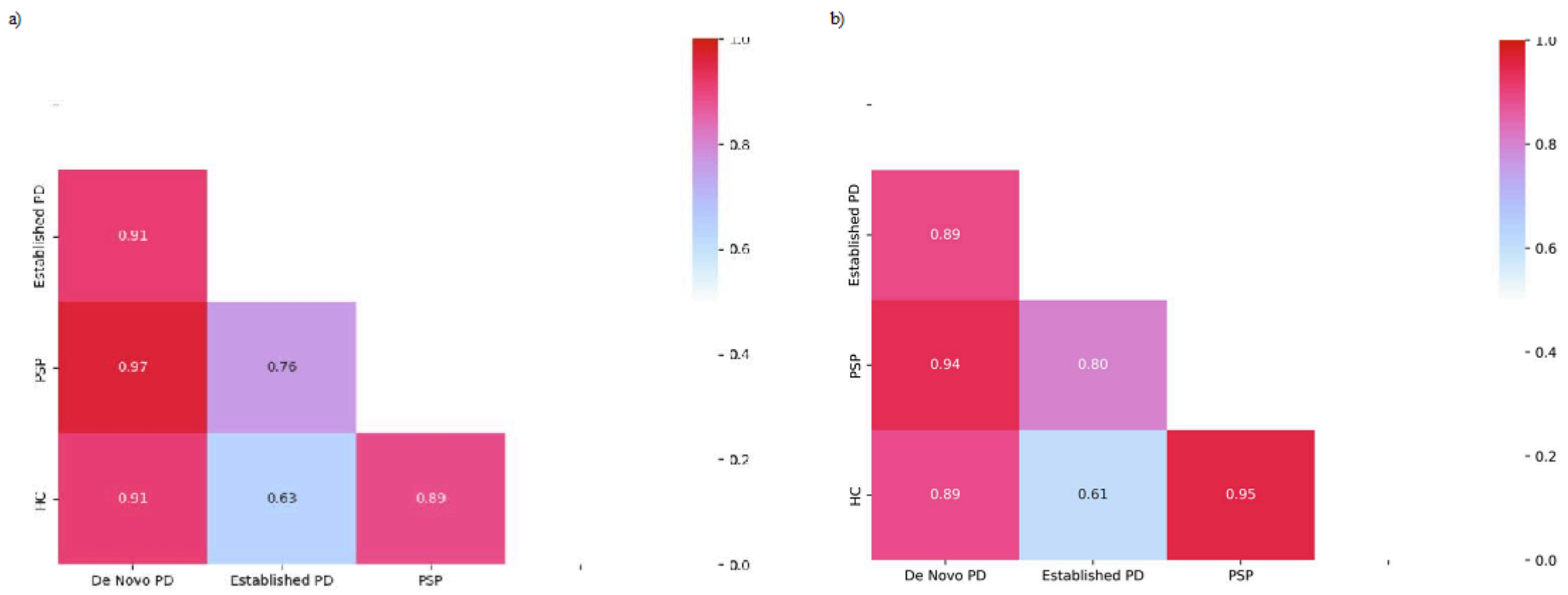

Supplementary Figure 1: *Cross validation methods using PCAxNB across all binary classification pairs and a) Stratified K Fold Cross Validation or b) Quicksort Leave One Pair Out Cross Validation*

3. **Performance Metrics for Model Pipelines**

In addition to the Area Under the Receiver Operating Characteristic (AUROC) presented in the main text, we evaluated several other performance metrics for each model pipeline and pairwise comparison. These metrics include Average Precision (AP), Youden J-statistic, Sensitivity, and Specificity. While the AUROC provides a comprehensive measure of a model's ability to discriminate between classes, these additional metrics offer complementary insights into the model's performance, particularly in the context of binary classification tasks.

Average Precision (AP) is a metric that summarises the precision-recall curve, which plots precision (positive predictive value) against recall (sensitivity) at different classification thresholds. AP is calculated by taking the weighted mean of precisions at each threshold, with the increase in recall from the previous threshold used as the weight. In our binary classification setting, AP quantifies the model's ability to correctly identify positive instances (e.g., de novo PD) while minimising false positives across various decision thresholds. A higher AP indicates better performance, with a perfect model achieving an AP of 1. AP is particularly informative when dealing with imbalanced datasets, as it emphasises the model's performance on the minority class.

Sensitivity, also known as recall or true positive rate, measures the proportion of actual positive cases that are correctly identified by the model. In our binary classification tasks, sensitivity represents the model's ability to correctly identify individuals belonging to a specific group (e.g., de novo PD) among all individuals who truly belong to that group. A high sensitivity indicates that the model effectively minimises false negatives, which is crucial when the cost of missing a positive instance is high.

Specificity, also known as true negative rate, measures the proportion of actual negative cases that are correctly identified by the model. In our study, specificity represents the model's ability to correctly identify individuals not belonging to a specific group (e.g., not de-novo PD) among all individuals who truly do not belong to that group. A high specificity indicates that the model effectively minimises false positives, which is important when the cost of misclassifying a negative instance is high.

The Youden J-statistic, is a measure of the overall diagnostic effectiveness of a binary classification model. It is calculated as the sum of sensitivity and specificity minus one (Sensitivity + Specificity - 1). The Youden J-statistic ranges from -1 to 1, with higher values indicating better diagnostic performance. A value of 1 represents a perfect test, while a value of 0 indicates a test with no discriminatory power. In our study, the Youden J-statistic provides a single metric that balances the model's ability to correctly identify positive instances (sensitivity) and negative instances (specificity) at the optimal classification threshold determined by maximising the index.

Table 1 presents these performance metrics for each model pipeline and pairwise comparison, providing a comprehensive evaluation of their diagnostic capabilities. By considering these metrics alongside the AUROC, we gain a more complete understanding of the strengths and limitations of each model in distinguishing between different participant groups.

| | **Pairwise Comparison** | **AU-ROC** | **Average Precision** | **Youden J-statistic** | **Sensitivity** | **Specificity** |
|---|---|---|---|---|---|---|
| Conventional RF+NB | De Novo PD vs Established PD | 0.45 | 0.17 | 0.16 | 0.27 | 0.88 |
| Conventional RF+NB | De Novo PD vs PSP | 0.75 | 0.79 | 0.54 | 0.64 | 0.90 |
| Conventional RF+NB | De Novo PD vs HC | 0.63 | 0.38 | 0.31 | 0.36 | 0.94 |
| Conventional RF+NB | Established PD vs PSP | 0.60 | 0.90 | 0.30 | 0.90 | 0.40 |
| Conventional RF+NB | Established PD vs HC | 0.64 | 0.78 | 0.28 | 0.65 | 0.63 |
| Conventional RF+NB | PSP vs HC | 0.75 | 0.53 | 0.69 | 0.80 | 0.89 |
| PCA+NB | De Novo PD vs Established PD | 0.89 | 0.59 | 0.72 | 1.00 | 0.72 |
| PCA+NB | De Novo PD vs PSP | 0.94 | 0.91 | 0.92 | 1.00 | 0.92 |
| PCA+NB | De Novo PD vs HC | 0.89 | 0.73 | 0.71 | 1.00 | 0.71 |
| PCA+NB | Established PD vs PSP | 0.80 | 0.92 | 0.62 | 0.87 | 0.75 |
| PCA+NB | Established PD vs HC | 0.61 | 0.76 | 0.29 | 0.55 | 0.74 |
| PCA+NB | PSP vs HC | 0.95 | 0.90 | 0.80 | 0.92 | 0.89 |
| PCA+MLP | De Novo PD vs Established PD | 0.79 | 0.56 | 0.55 | 0.73 | 0.83 |
| PCA+MLP | De Novo PD vs PSP | 0.97 | 0.97 | 0.83 | 1.00 | 0.83 |
| PCA+MLP | De Novo PD vs HC | 0.72 | 0.45 | 0.39 | 0.91 | 0.49 |
| PCA+MLP | Established PD vs PSP | 0.86 | 0.96 | 0.72 | 0.97 | 0.75 |
| PCA+MLP | Established PD vs HC | 0.65 | 0.75 | 0.28 | 0.83 | 0.46 |
| PCA+MLP | PSP vs HC | 0.91 | 0.82 | 0.72 | 0.92 | 0.80 |
| Sym_AE+NB | De Novo PD vs Established PD | 0.81 | 0.37 | 0.58 | 0.91 | 0.67 |
| Sym_AE+NB | De Novo PD vs PSP | 0.92 | 0.85 | 0.92 | 1.00 | 0.92 |
| Sym_AE+NB | De Novo PD vs HC | 0.58 | 0.3 | 0.37 | 0.91 | 0.46 |
| Sym_AE+NB | Established PD vs PSP | 0.85 | 0.96 | 0.67 | 0.84 | 0.83 |
| Sym_AE+NB | Established PD vs HC | 0.66 | 0.8 | 0.36 | 0.51 | 0.86 |
| Sym_AE+NB | PSP vs HC | 0.90 | 0.68 | 0.77 | 1.00 | 0.77 |
| Sym_AE+MLP | De Novo PD vs Established PD | 0.79 | 0.33 | 0.54 | 0.73 | 0.81 |
| Sym_AE+MLP | De Novo PD vs PSP | 0.93 | 0.93 | 0.75 | 1.00 | 0.75 |
| Sym_AE+MLP | De Novo PD vs HC | 0.62 | 0.41 | 0.28 | 0.91 | 0.37 |
| Sym_AE+MLP | Established PD vs PSP | 0.82 | 0.95 | 0.59 | 0.84 | 0.75 |
| Sym_AE+MLP | Established PD vs HC | 0.53 | 0.71 | 0.12 | 0.38 | 0.74 |
| Sym_AE+MLP | PSP vs HC | 0.92 | 0.82 | 0.75 | 0.92 | 0.83 |

*Supplementary Table 1: Performance metrics for each model pipeline and pairwise comparison between participant groups. Metrics include area under the receiver operating characteristic curve (AUROC), average precision (AP), sensitivity, and specificity. All metrics range from 0 to 1, with higher values indicating better performance. The Youden J-statistic, used to determine the optimal sensitivity and specificity for each model, ranges from -1 to 1, with 1 representing a perfect test and 0 indicating no discriminatory power. AUROC and AP were calculated using quicksort leave-pair-out cross-validation, while sensitivity and specificity were derived from the point on the ROC curve that maximized the Youden J-statistic.*

**4. Normalisation Comparison**

To assess the impact of different normalisation methods for the waveform reconstruction approach, we compared the performance of our best-performing hierarchical model, PCAxNB, using two distinct preprocessing techniques: position input, and velocity input. Position input aligns the waveforms based on their minimum value and normalises them according to their rightmost (last) value. The velocity input introduces an additional step of computing the gradient of the position normalised waveforms, effectively capturing the instantaneous rate of change in eye position. Gradients were calculated using NumPy's default implementation, by second-order central differences except at the first and last time points, where first-order one-sided differences were used.

Across the waveform reconstruction models, the velocity inputs consistently outperformed position inputs. This suggests that incorporating the velocity information, which encodes the dynamic characteristics of the saccadic eye movements, provides supplementary discriminative features that enhance the model's ability to differentiate between the groups. The superior performance of velocity input underscores the significance of considering not only the static attributes of the saccade waveforms, such as their shape and relative position, but also the temporal dynamics when analysing saccadic eye movements for diagnostic purposes.

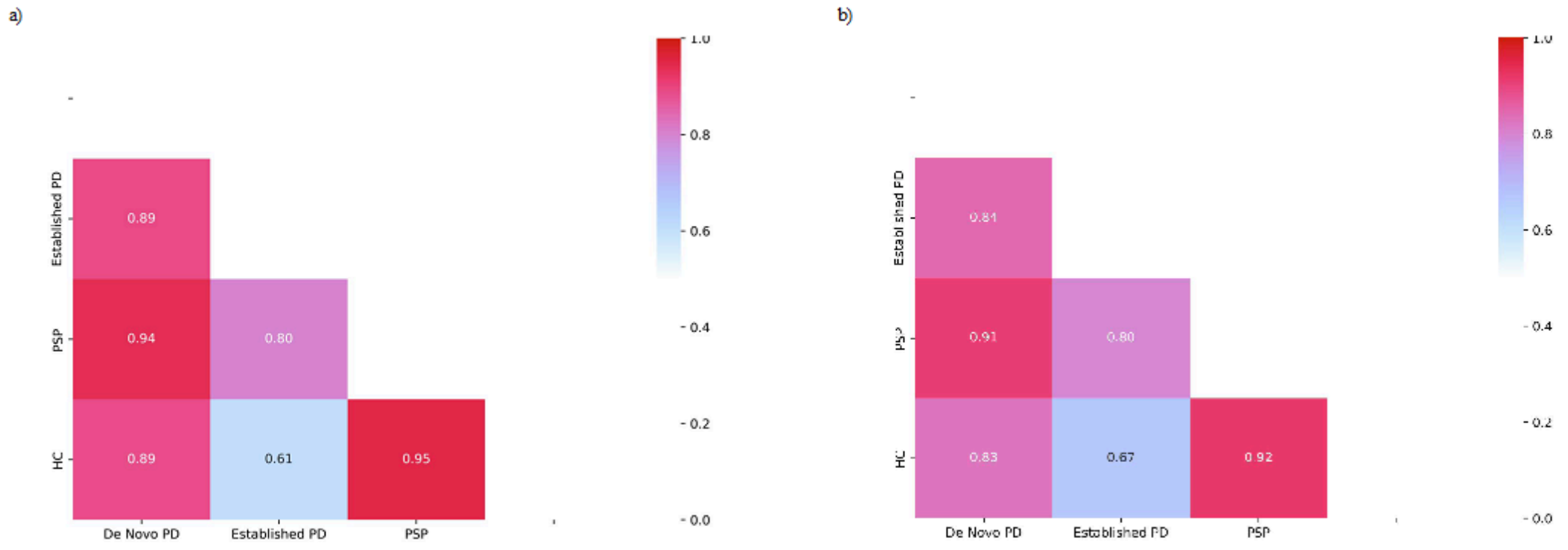

Supplementary Figure 2: *Normalisation methods applied to the raw saccade waveforms before classifying with our best performing model. a) Velocity input, b)* Position *input*

#### 4. Down-sampling Analysis

To assess the robustness of our hierarchical waveform reconstruction approach to reduced temporal resolution, we conducted a down-sampling analysis focusing on our best-performing model, PCAxNB. The raw saccade waveforms, originally sampled at 200 Hz, were down-sampled to 100 Hz and 50 Hz using the decimate function from SciPy. The decimate function applies an anti-aliasing filter before down-sampling to prevent aliasing artifacts, ensuring that the down-sampled waveforms accurately represent the original data.

The choice of 100 Hz and 50 Hz as down-sampling rates was based on their relevance to real-world scenarios, as these sampling rates are commonly found in lower-frequency eye-tracking devices such as phone cameras or budget-friendly eye trackers. By evaluating the performance of our approach at these reduced sampling rates, we can assess its applicability to a wider range of data acquisition setups.

The down-sampled datasets were then processed using our hierarchical classification pipeline, with performance evaluated using pairwise AUROC scores. This analysis serves two purposes: a) Investigate the impact of reduced sampling rates on classification accuracy, simulating scenarios with different eye-tracking devices. b) Identify the minimum sampling rate required to maintain satisfactory performance, guiding device selection and data acquisition protocols.

As shown in Figure S4, the classification performance of the PCAxNB model had a noticeable decline in performance when the sampling rate was reduced to 100 Hz and 50 Hz. These results suggest that a minimum sampling rate of 200 Hz is required to maintain sufficient classification accuracy across participant groups.

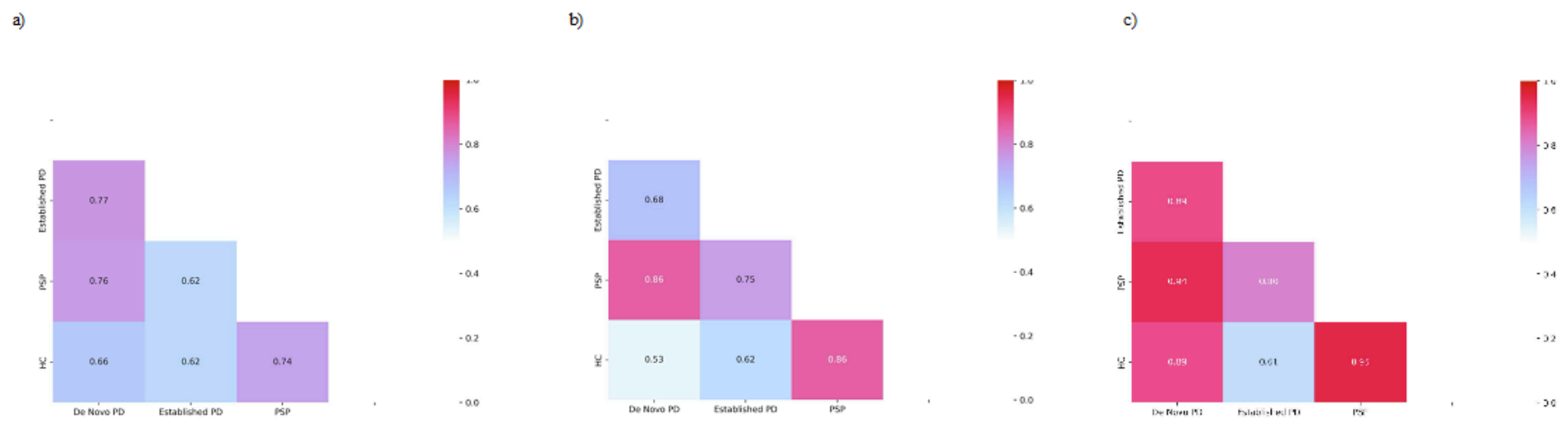

Supplementary Figure 3: *Heatmaps depicting the pairwise classification performance (AUROC) of the PCAxNB model on down-sampled datasets. a) Down-sampled to 50 Hz, b) Down-sampled to 100 Hz, c) Original 200 Hz dataset without down-sampling.*